\documentclass[preprint,12pt]{elsarticle}

\usepackage{xcolor}

\definecolor{custom-gray}{cmyk}{0, 0, 0, 0.7, 1.00}
\usepackage[most]{tcolorbox}
\newtcbtheorem[no counter]{Summary}{\hskip-0.97em}{enhanced,drop shadow={black!50!white},
 coltitle=white,
 top=0.15in,
 separator sign none,
 attach boxed title to top left=
 {xshift=0em,yshift=-\tcboxedtitleheight/2},
 boxed title style={size=small,colback=custom-gray}
}{summary}

\usepackage{xurl}
\usepackage{hyperref}

\usepackage{threeparttable}

\usepackage{booktabs}
\usepackage{multirow}

\newcommand{\responsecolorbox}[2][]{%
  \colorbox{#1}{\small \parbox{\dimexpr\linewidth-2\fboxsep}{#2}}
}

\def\inline{\lstinline[basicstyle=\ttfamily \footnotesize]}

\newtcolorbox{siderules}{
  breakable,
  enhanced jigsaw,
  oversize,
  rightrule=0pt,
  toprule=0pt,
  bottomrule=0pt,
  colback=white,
  arc=0pt,
  outer arc=0pt,
  title style={white},
  fonttitle=\color{black}\bfseries,
  top=0pt,
  bottom=0pt,
  left=1pt,
}

\usepackage{scalerel}
\usepackage{tikz}
\usetikzlibrary{svg.path}

\definecolor{orcidlogocol}{HTML}{A6CE39}
\tikzset{
  orcidlogo/.pic={
    \fill[orcidlogocol] svg{M256,128c0,70.7-57.3,128-128,128C57.3,256,0,198.7,0,128C0,57.3,57.3,0,128,0C198.7,0,256,57.3,256,128z};
    \fill[white] svg{M86.3,186.2H70.9V79.1h15.4v48.4V186.2z}
                 svg{M108.9,79.1h41.6c39.6,0,57,28.3,57,53.6c0,27.5-21.5,53.6-56.8,53.6h-41.8V79.1z M124.3,172.4h24.5c34.9,0,42.9-26.5,42.9-39.7c0-21.5-13.7-39.7-43.7-39.7h-23.7V172.4z}
                 svg{M88.7,56.8c0,5.5-4.5,10.1-10.1,10.1c-5.6,0-10.1-4.6-10.1-10.1c0-5.6,4.5-10.1,10.1-10.1C84.2,46.7,88.7,51.3,88.7,56.8z};
  }
}

\newcommand\orcidicon[1]{\href{https://orcid.org/#1}{\mbox{\scalerel*{
\begin{tikzpicture}[yscale=-1,transform shape]
\pic{orcidlogo};
\end{tikzpicture}
}{|}}}}

\usepackage{hyperref}

\usepackage{etoolbox}

\usepackage{natbib}

\usepackage{subfig}

\begin{document}

\begin{frontmatter}

\title{Is Your Private Information Logged? An Empirical Study on Android App Logs}



\author[inst1]{Zhiyuan Chen}
\author[inst1]{Soham Sanjay Deo}
\author[inst1]{Poorna Chander Reddy Puttaparthi}
\author[inst1]{Vanessa Nava-Camal}
\author[inst1]{Yiming Tang}
\author[inst1]{Xueling Zhang}
\author[inst2]{Weiyi Shang}

\address[inst1]{Department of Software Engineering, Rochester Institute of Technology, NY, USA}
\address[inst2]{Department of Electrical and Computer Engineering, University of Waterloo, ON, Canada}

\date{Received: date / Accepted: date}







\begin{abstract}
    With the rapid growth of mobile apps, users' concerns about their privacy have become increasingly prominent. Android app logs serve as crucial computer resources, aiding developers in debugging and monitoring the status of Android apps, while also containing a wealth of software system information. Previous studies have acknowledged privacy leaks in software logs and Android apps as significant issues without providing a comprehensive view of the privacy leaks in Android app logs. In this study, we build a comprehensive dataset of Android app logs and conduct an empirical study to analyze the status and severity of privacy leaks in Android app logs. Our study comprises three aspects: (1) Understanding real-world developers' concerns regarding privacy issues related to software logs; (2) Studying privacy leaks in the Android app logs; (3) Investigating the characteristics of privacy-leaking Android app logs and analyzing the reasons behind them. Our study reveals five different categories of concerns from real-world developers regarding privacy issues related to software logs and the prevalence of privacy leaks in Android app logs, with the majority stemming from developers' unawareness of such leaks. Additionally, our study provides developers with suggestions to safeguard their privacy from being logged.
\end{abstract}

\begin{keyword}
privacy leak \sep android, logging \sep software logs
\end{keyword}

\end{frontmatter}

\section{Introduction}

In today's fast-paced world, mobile applications (apps) have become an integral part of our daily lives, aiding us in completing numerous essential activities and tasks~\cite{Amalfitano2013, Hilmi2021}. From facilitating social connections to providing entertainment, these apps play a pivotal role in our lives. Mobile apps often process sensitive user data, such as names, email, and phone numbers. To protect user's privacy, several privacy-related policies such as the California Consumer Privacy Act (CCPA)  \cite{ccpa} and Children's Online Privacy Protection Act (COPPA) \cite{coppa} emphasize the importance of protecting personal data and set strict guidelines for its collection, processing, and storage. Violations of these regulations can lead to legal consequences and substantial financial penalties.  For example, the Federal Trade Commission (FTC) fined Facebook \cite{facebook} for sharing user data with third parties, and a fertility app faced charges for health data sharing without user notification in 2023~\cite{fertility}.

Given that the Android OS is currently the most popular mobile operating system in the global market, boasting the largest user base~\cite{AppMySite2024}, the studies of privacy leaks in Android apps have gained significant attention. The existing research usually focuses on network~\cite{Song2015, Chen2015} and program analysis~\cite{Chen2023, Yang2013, Jain2017}, encompassing both static and dynamic approaches. These methodologies primarily scrutinize data transmissions and code behaviors to uncover potential privacy violations. 

Similar to any other software, Android apps also generate logs that are an essential resource used to record systems' runtime information and play a significant role in numerous software development activities, such as debugging~\cite{Li2018, Chen2017, Yuan2012, Zhu2015} and program comprehension~\cite{Chen2017, Fu2013}. Through such recording, private information may be proactively or accidentally captured by logging while later leaked when these log data is analyzed or archived in various software engineering activities.


The risk of user information leakage in Android logs is increasingly critical.  
First, Android apps are widely used by end-users on personal devices for daily activities and often have access to large amounts of sensitive data.  Second, Android apps are typically not standalone software and usually require a connection to a network and communication with a remote server. Data transmission through the network can potentially lead to the exposure of users' private information to remote servers or unauthorized third parties, resulting in unforeseeable privacy leaks. The connectives also enable the transmission of logs that may carry sensitive information to be collected over the network. 
Third, access to Android app logs is not strictly restricted. The Android official documentation~\cite{Log_Google_2024} notes that although the Android system restricts permissions to access the log memory buffer \inline{logcat}, there are still apps that can access it as Android needs to support a wide variety of devices.
Finally, Android app logs are not necessarily destroyed after the apps' execution. The logs can persist in the form of files on the mobile devices. Therefore, privacy leaks do not necessarily conclude with the end of app execution and may persist as a long-term issue, especially since log files typically lack access control protection.
The following Android app log serves as an example of privacy leaks, as it records sensitive user information (a user name). 
To protect user privacy, we erased the value of the user name.
Due to the easily accessible and long-term nature of Android app logs discussed above, users' sensitive information may be leaked to the public, leading to adverse consequences.


\noindent
\responsecolorbox[gray!10]{
 09-26 15:54:06.054 20811 20811 D m : cloud\_user\_info\_name = (string) {\color{red} ...}
}

To better protect users' privacy, the Android's official documentation~\cite{Android_doc_2024} emphasizes the importance of security and privacy in Android app logs. The Android's official document \cite{androidLogRequirement} also warns developers not to include sensitive information in logs:

\begin{siderules}
\emph{``Do not log Personally Identifiable Information (PII) such as Email addresses, telephone numbers and names. Similarly, certain details are considered sensitive even if not explicitly personally identifiable.''}
\end{siderules}

Studying and mitigating privacy leak issues in Android app logs is crucial, 
yet research in this area remains limited. The most related work~\cite{Lyons2023} focuses on privacy leak status in Android app logs but fails to incorporate insights from real-world developers or conduct deeper analysis, such as examining privacy leaking issues using a logging research perspective to reveal the characteristics of privacy-leaking logs.
To address this research gap, we conducted an empirical study on Android app logs through three distinct dimensions: (1) collecting and summarizing real-world developers' concerns about privacy leak issues related to software logs, (2) studying privacy leaks in a comprehensive dataset of Android app logs, and (3) analyzing the characteristics of privacy-leaking Android app logs. Our study identified five different categories of developers' concerns regarding privacy issues related to software logs and found that developers' concerns are indeed warranted based on the privacy leaks we detected in Android app logs. We identified 610 instances of privacy leaks in the log files of 83 Android apps, with 21 instances originating from third-party libraries, indicating common occurrences of privacy leaks in Android logs and confirming that users' privacy is transmitted remotely. Through our analysis of privacy-leaking logs, we observed that a significant portion of the logs have a high logging level, implying that they are printed during runtime rather than omitted. Moreover, developers are often unaware of privacy leaks in Android app logs, highlighting the necessity for developers to enhance their privacy awareness while logging for Android apps.

The contributions of this paper are as follows:

\begin{itemize}
    \item  We have constructed a log dataset of Android apps, which encompasses 83 diverse and popular Android apps, thereby addressing the existing gaps in currently available open-source Android app logs. 
    \item We detected non-trivial privacy-leaking instances from Android app logs, indicating that this issue should be given attention by developers.
    \item This is the first empirical study to investigate the characteristics of privacy-leaking Android app logs, explore the root causes of privacy leaks, and propose corresponding suggestions.
    \item To facilitate open-source community and academic communication, we have made our code and data publicly available\footnote{\url{https://github.com/android-app-logs/Android-App-Log-Privacy-Study}}.
\end{itemize}

\textbf{Paper Organization.} 
The remaining sections of the paper are structured as follows. 
Section~\ref{sec:back} presents the study background and related work.
Section~\ref{sec:study} presents our study design. 
Section~\ref{sec:experiment} presents the experiments, which include how we addressed the three RQs and gained results and findings.
Section~\ref{sec:discussion} provides additional discussions on the results of the RQs.
Section~\ref{sec:threat} explores potential threats to validity.
Section~\ref{sec:conc} concludes the paper and outlines directions for future work.

\section{Background and Related Work}~\label{sec:back}

In this section, we present the background and related work of our study.

\subsection{Privacy Leaks in Software Logs}

Software logs are utilized to record software runtime information, and their significance is widely acknowledged in many existing studies~\cite{Ding2024, Ding2023, Zhang2022, He2016}.
As modern software systems become increasingly complex and highly transactional in nature~\cite{Tang2022}, the size of logs grows rapidly, and the logs contain vast amounts of information, which makes it difficult for developers to manually inspect and analyze them. Therefore, if software logs include user privacy, this issue may also be overlooked by the developers, leading to privacy leaks.
For example, 
\citet{Christof2016} highlights the threats posed by unauthorized access to log records, thus proposing a framework for generating log records that do not contain sensitive information.
\citet{Licorish2015} conduct an empirical study on the logs of three different Android OS releases, which reveal varying degrees of severity in terms of security and privacy concerns for each release.
\citet{Ito2018} mention that the misuse of privacy information by authorized apps is becoming a serious issue, and they address this by detecting privacy leaks through the analysis of API call logs.
\citet{Agrawal2019} propose a distributed log processing algorithm that utilizes Locality Sensitive Hashing (LSH) to encrypt log lines.
\citet{Stephan2019} discover that event logs from information systems may contain sensitive information about employees, and they propose an approach to sanitize logs to protect employee privacy.
Despite the abundance of research on privacy leaks in software logs, to the best of our knowledge, there is a limited empirical study focused on investigating privacy leaks in Android app logs. The most related work~\cite{Lyons2023} also conducts an empirical study on Android app logs, but our work differs from theirs in the following aspects: (1) Our privacy leakage detection is 100\% accurate as the search for privacy information is based on strict string matching without fuzzy search. (2) In addition to investigating privacy leakage in Android app logs, we also explored real-world developers' perspectives on Android app log leaks. (3) We are the first study to examine the characteristics of privacy-leaking Android app logs. Their paper is published in a security venue, and it lacks analysis of privacy leaks in Android app logs from a logging research perspective. (4) We have identified the status of privacy leaks originating from third-party libraries. 


\subsection{Privacy Leaks in Android Apps}

In modern society, mobile apps play a crucial role in many daily activities, such as communication, business, entertainment, etc.~\cite{Amalfitano2013, Hilmi2021}.
According to existing statistics, Android is the most popular mobile OS in the global market~\cite{Josh2024} and attracts billions of users~\cite{AppMySite2024}. Therefore, studying privacy protection and leaks in Android apps has the potential to yield a broad impact in practice. 

Many previous studies have demonstrated that privacy leaks in Android apps are a significant issue and have proposed numerous approaches to enhance privacy protection and raise developers' awareness of privacy concerns. There has been a lot of work on the detection of information leaks on mobile apps based on statistical or dynamic program analysis. In particular, 
\textsc{FlowDroid}~\cite{arzt2014flowdroid} leverages static taint analysis to trace information for detecting information leaks. 
\textsc{TaintDroid}~\cite{enck2014taintdroid}  dynamically observes how sensitive information is used and transmitted by tracking apps' data flow and reveals that a significant portion of Android apps introduce potential privacy risks to sensitive user data.
\textsc{VetDroid}~\cite{zhang2013vetting} is a dynamic analysis platform designed to capture permission usage behavior in real time by intercepting Android API calls to identify and analyze potential information leaks.
Han et al. \cite{han2012study}  employ dynamic taint analysis to investigate the exposure of personal data and persistent identifiers in information flow within apps. 
\textsc{ConDySTA} \cite{zhang2021condysta} incorporates dynamic taint analysis results to complement static taint analysis while maintaining context sensitivity, and they successfully identify numerous data leaks within real-world Android apps.
\citet{Slavin2016} propose a framework to verify the consistency between Android apps' privacy policies and their code, which conducts information flow analysis to identify misalignments between policy statements and actual code implementation.
\citet{Yang2013} propose an analytical framework using symbolic execution to help users distinguish between Android apps that lead to privacy leaks and secure apps.
\citet{Chen2023} propose a combined approach using static and dynamic analysis to detect privacy leaks in massive Android apps.
\citet{Gibler2012} conduct an empirical study on $\sim$24K Android apps to uncover privacy leak issues, and $\sim$2K apps were manually verified to have leaked personal data.
\citet{Jain2017} propose an automaton-based static analysis framework for detecting privacy leaks of Android inter-apps.
\citet{Li2015} propose a static taint analyzer to detect privacy leaks among Android app components.

Various techniques in network traffic analysis have also been employed to identify personal data shared by Android apps with third parties. Specifically, Razaghpanah et al. \cite{razaghpanah2018apps}  focus on identifying third-party advertising and analytics services, while Ren et al. \cite{ren2016recon} instrument VPN servers to uncover privacy leaks in network traffic. Additionally, Vallina et al. \cite{vallina2012breaking} analyze mobile ISP traffic logs to pinpoint advertisement traffic. \citet{Song2015} conduct a study to examine network traffic from Android apps to detect privacy leaks.
These prior methodologies primarily scrutinize data transmissions and code behaviors to uncover potential privacy violations.


While the aforementioned studies attempt to address privacy leaking issues within Android apps from different perspectives, there is currently a lack of disclosure and solutions for privacy leaks in Android app logs from a logging research perspective.

\subsection{Privacy Awareness about Information Leaks}
Privacy awareness refers to users' understanding of their privacy status concerning disclosure and protection~\cite{Malandrino2013}. 
While disclosing users' privacy to the public is a serious issue, the situation becomes even more severe when privacy is leaked without users' awareness. The level of users' privacy awareness could serve as an indicator to assess the severity of privacy leaks. Many existing studies are dedicated to enhancing users' privacy awareness.
For example, 
\citet{Stefanie2008} explore addressing the privacy paradox, where users possess privacy awareness but do not behave according to their stated attitudes.
\citet{Malandrino2013} develop a comprehensive and efficient client-side tool to maximize users' awareness of the extent to which their information is being leaked.
\citet{Hazari2013} conduct a study to investigate users' attitudes toward privacy on social networking sites.
\citet{Bergmann2008} introduce a test to empirically measure changes in user privacy awareness. 
These studies all indicate that there is room for improvement in users' privacy awareness. However, none of these studies have focused on users' privacy awareness in Android app logs, which necessitates further investigation.


While previous research has made progress in uncovering privacy issues in general software logs and Android apps and in enhancing user privacy awareness, the privacy issues within Android app logs have remained unexplored. Our work addresses this research gap by not only detecting privacy leaks within Android logs but also analyzing the characteristics of logs that lead to privacy leaks. By identifying root causes, we aim to assist developers in raising awareness about privacy protection and effectively mitigating risks. Given the immense scope of the Android market, our work is poised to have a widespread and tangible impact on the safeguarding of personal information in the digital age.

\section{Study Design}\label{sec:study}

In this section, we present the design of our research questions and the methodology for collecting data for our experimental setup.









\begin{figure}[h]
    \centering
    \includegraphics[width=0.65\textwidth]{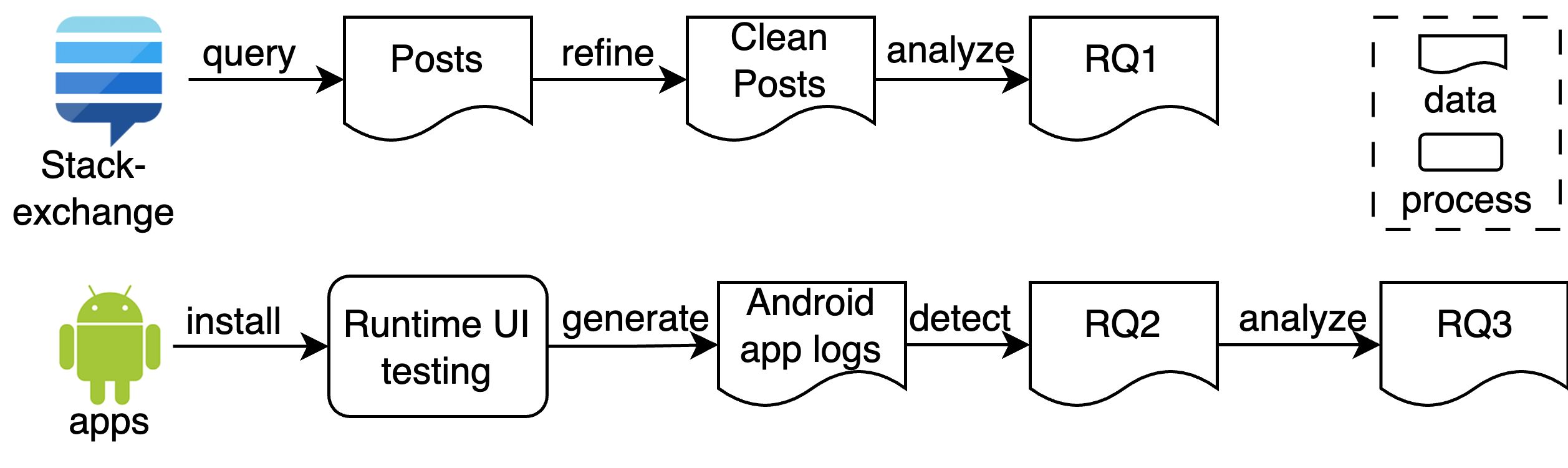}
	\caption{An overview of our study. } 
	\label{fig:workflow} 
 \end{figure}


\subsection{Research Questions}
This study consists of three research questions. Figure~\ref{fig:workflow} illustrates an overview of our study. RQ1 is an independent study, while RQ2 aims to detect privacy leaks in Android app logs. RQ3 builds upon the results of RQ2 and conducts further analysis of privacy leak cases.

\noindent \textit{\textbf{\underline{RQ1:}} In real-world software development, what privacy concerns do developers have related to software logs?}

The motivation behind this research question stems from the growing importance of privacy concerns in software development, particularly given the potential for privacy leaks through software logs, as mentioned in Section~\ref{sec:back}. As modern software systems become increasingly complex, developers are faced with the task of managing vast amounts of data related to program activities, such as software logs. In handling such large datasets, developers may inadvertently overlook privacy issues. Therefore, developers should strike a delicate balance between data management and respecting user privacy. By exploring privacy issues related to software logs encountered by developers, it is possible to implement more privacy-respecting logging practices in real-world software development scenarios. 

\noindent \textit{\textbf{\underline{RQ2:}} Do Android app logs contain privacy leaks?}

This research question is motivated by the concerns of privacy leaks in Android apps. With the pervasive use of the Android OS and the extensive logging practices employed by Android apps, there is a pressing need to investigate whether Android app logs pose any risks to user privacy. This RQ aims to uncover the current status of privacy leaks within Android app logs.
Moreover, security and privacy protection are required by the Android official documents~\cite{Android_doc_2024, androidLogRequirement}. This RQ also intends to see whether developers follow these guidelines in their apps.

\noindent \textit{\textbf{\underline{RQ3:}} What are the characteristics of Android app logs that leak private information? }

The research question is a further study based on RQ2. Given the privacy leaking cases in Android app logs, we aim to understand the severity of privacy leaking issues by identifying and analyzing the characteristics of the logs, and to gain insight to avoid privacy leaks in Android app logs.




\subsection{Data Collection: Android App Log Generation and Collection.}

The data for this study is derived from two main sources: (1) querying real-world Q\&A developer forums to gather real-world developers' concerns regarding privacy issues related to software logs (RQ1). The relevant details can be found in the approach section of RQ1; and (2) a large dataset of Android app logs (RQ2 \& RQ3). The construction of this log dataset involves the following three steps:

\paragraph{Subject apps}
We collected our subject apps from PlayDrone~\cite{metadata}, a collection of metadata for Android apps on the Google Play store, which has been examined in many existing studies~\cite{Ficco2019, Angelo2020, Xueling2023}.
Specifically, we selected the top 100 Android apps and finally successfully installed and executed 83 apps on our test device.
Since PlayDrone was collected some time ago, we additionally attempted to download the latest versions of these 83 apps and ultimately obtained Android app logs for 78 apps. Several apps could not be included for the following reasons: Dianxinos, Walmart Grocery, and UCMobile were unavailable in our region due to geo-restrictions on the Play Store; BBM blocked the automated testing tool (Monkey), preventing proper execution; and the APK package com.abtnprojects.ambatana could no longer be found on the Play Store.

 \paragraph{User profile for testing}
We set up an Android device and collected its information to construct a reference user profile. The profile includes different platform IDs (e.g., \textit{device ID}, \textit{serial number}, \textit{Android ID}, \textit{advertising ID}) and a synchronized Google account (e.g., \textit{user name}, \textit{full name}, \textit{user email}). 
The list of values in the user profile will be presented in detail in Table~\ref{tab:table_rq2_info}, but we omit some characters for privacy protection purposes.


\paragraph{Log generation}
We generate logs by performing  UI testing on the apps under study. For each app, we use the Android Debug Bridge (adb)~\cite{Google-adb} to automatically install the apps onto our test device and run \textsc{Monkey}~\cite{developers2012ui} to conduct the testing. \textsc{Monkey} is a tool developed as a part of the Android toolkit to perform GUI testing on apps. The tool generates different types of UI events that can interact randomly with the activity components of an app, such as clicks, drags, and touches. We use \textsc{Monkey} to perform testing because it is the most well-established event generator used in Android dynamic analysis, and it is fully automated and robust enough to be applied to all apps. Furthermore, existing studies~\cite{wang2018empirical} show that \textsc{Monkey} achieves comparable coverage with state-of-the-art tools. After testing, we automatically save the system log into the local file system for later inspection. For apps that require registration and login during testing, we manually create accounts using the user profile data we predefined in Table~\ref{tab:table_rq2_info} to complete the login process. 
Below is a log example from the Discord Android app, which is used to record information about HTTP requests. 
In this example, there are three log lines, each containing two different types of content: log messages and the other. Log messages refer to the content following the colon, which includes the information developers want to log. The information before the colon is the metadata of the logs, most of which are automatically generated by the logging framework.
In the first log line, the timestamp is \inline{09-26 15:29:56.542}, the process ID is \inline{18466}, the thread ID is \inline{18548}, the logging level is Verbose (\inline{V}), and the tag is \inline{Discord}. The log message is: \inline{--\textgreater~GET https://discordapp.com/api/v6/auth/}
\inline{consent-required}. Log messages are the focus of our study, where private information could be present.


\noindent
\responsecolorbox[gray!10]{
09-26 15:29:56.542 18466 18548 V Discord : --\textgreater~GET https://discordapp.com/api/v6/auth/consent-required\\
09-26 15:29:56.584 18466 18548 V Discord : Breadcrumb[http]: HTTP[GET] - https://discordapp.com/api/v6/auth/consent-required\\
09-26 15:29:56.602 18466 18538 V Discord : \textless-- 200 https://dl.discordapp.net/apps/android/versions.json (3408ms, 42-byte body)
}



In the end, we successfully collected 83 Android app log files with a total of 667,702 log lines, as well as 78 log files corresponding to the latest versions of these Android apps, with a total of 8,671,009 log lines.

\section{Experimental Results}\label{sec:experiment}
In this section, we present the results of our study. 

\subsection{Developers' Concerns about Privacy Issues Related to Software Logs (RQ1).}


\noindent \textbf{\underline{Approach.}} 
To answer this question, we explored real-world discussions on developers' Q\&A forums about privacy leaking issues related to software logs. We use the Stack Exchange Data Explorer API\footnote{\url{https://api.stackexchange.com/docs}} to search for privacy issues related to software logs. Stack Exchange stores all user-contributed content on its network, including Stack Overflow~\footnote{\url{https://stackoverflow.com/}}, the largest Q\&A platform for developers. Stack Exchange provides an API for users to write SQL queries to search for Q\&A data. In this RQ, we use the keywords \inline{log} and privacy-related terms such as \inline{privacy}, \inline{sensitive information}, \inline{data leak}, etc., for search posts. 
The full list of keywords can be found in our project repository~\cite{Chen2024}.
We also exclude some keywords to avoid unexpected query results, such as \inline{log in}, \inline{log out}, \inline{logic}, etc. To obtain more accurate results,
we further conduct manual analysis to filter out the unexpected results. We manually label each post to determine if it is related to logs and privacy.
Two participants independently label each post. If a disagreement cannot be resolved through discussion, a mediator is engaged as a third participant until a consensus is reached.

\noindent  \textbf{\underline{Results.}}
After the search, we found 616 posts. Following manual analysis, a total of 59 posts mentioning software logs were identified as relevant to user privacy. 
These 59 posts were further categorized into five categories.
Participants first identified a few initial categories by checking some posts and reaching a consensus. During the subsequent labeling process, participants used the initial categories to label the posts. If they encountered a post that did not fit any existing category, they marked it and discussed adding a new category.

Below is an introduction to the five categories:

\paragraph{(a) The developers are attempting to replace sensitive user information in the software logs with non-sensitive information (35.59\%)} The replacement could be done using wildcards, total deletion, template strings, etc.
For example, the developer proposed a replacement strategy:

\begin{siderules}
\textit{
I need a way to replace "access\_token" and "client\_secret" within a string with [FILTERED]
}
\end{siderules}

In the entire post above, the developer mentioned HTTP request headers, request bodies, and responses might carry sensitive information. This insight also highlights for developers the importance of considering privacy protection when logging information related to HTTP requests/responses.

Additionally, we also learned from a developers' post (see the example below) that JSON data structures might carry sensitive information, causing concern for the developers. 

\begin{siderules}
\textit{
I have some JSON input, the shape of which I cannot predict, and I have to make some transformations (to call it something) so that some fields are not logged.
}
\end{siderules}

\paragraph{(b) The developers use software logs to seek the solution to privacy-related issues (27.12\%)} In this category, developers intend to utilize information from logs to address privacy-related issues that are not specific to software logs themselves. In the following example, the developer encountered a privacy issue, but the developer did not understand it until he/she analyzed the logs.

\begin{siderules}
\textit{
I have a code which shares a line of text and an image via UIActivityViewController. After I updated to iOS 10 it started crashing with the following log: 
"This app has crashed because it attempted to access privacy-sensitive data without a usage description.  The app's Info.plist must contain an NSPhotoLibraryUsageDescription key with a string value explaining to the user how the app uses this data."
}
\end{siderules}

\paragraph{(c) The developers raise concerns about potential privacy leaks in software logs and are seeking help and suggestions on a Q\&A forum (27.12\%)} 
For example, the developer below wants to develop a Rails application and has a concern that the logs may store sensitive information.

\begin{siderules}
\textit{
We don't want sensitive information to be printed in logs.
We should be able to prevent reverse engineering the sensitive data in case the data is stolen.
}
\end{siderules}

\paragraph{(d) The developers do not want to protect privacy (6.78\%)} In this category, the developers do not want to protect privacy. They either aim to extract sensitive information from the logs or intend to leak privacy on purpose.
For example, the developer wants to build an APP for the iPhone that can extract call history from the logs:

\begin{siderules}
\textit{
 I want to make an app, wherein I have access to the following\\
• Logs: call log, message log, app usage log, media log, ...\\
• Logging data per activity: ex) call log: call direction, phone number, call
duration, location, ...\\
• Privacy policy (logging policy, using policy of the log data, user
consent policy, ...)
}
\end{siderules}


Below is another example where the developer wants to print sensitive information in the logs.

\begin{siderules}
\textit{
Can log4j be set to print my db uname\&pass, or any of the sensitive data used in my Hibernate config in the App???
}
\end{siderules}

\paragraph{(e) Protect privacy by log storage (3.39\%)} This could be implemented by carefully considering how to store the logs. Different storage methods and the strictness of access control will affect the ease of privacy leakage. For example, in the following post, privacy is protected by managing the access permissions of different sheets used to store logs.

\begin{siderules}
\textit{ I am working on a google spreadsheet for my team members where they can save their logs so I have made a separate sheet for each of them but I do not want to let them see each others logs due to some privacy reasons.\\
I am hiding sheets with the app script and associated them with the email address of the users. But  when any user opens his/her sheet then it works just fine, but if other users open their sheets then it get messed up.
}
\end{siderules}


Based on the categories above, we find that  
although most of the developers want to protect logs from leaking privacy (categories a, c and e), a portion of developers intend to do the opposite (category d). Additionally, while logs can result in privacy leaks, they are also considered an important vehicle for solving privacy issues (category b).


\vspace{0.1cm}
\begin{Summary}{Summary of RQ1}{}
Real-world developers have a variety of concerns regarding privacy issues related to software logs, including sensitive information replacement, the use of software logs, concerns about the log content, log storage, etc. The most common concern revolves around the strategies for replacing sensitive information in the logs.
\end{Summary}
\vspace{0.1cm}

\subsection{Studying Privacy Leaks in Android App Logs (RQ2).}

\noindent \textbf{\underline{Approach.}} To examine whether Android apps' logs contain privacy leaks, we first collect Android app log files from our Android device. In these logs, we search for personal information keywords specific to our sensitive information. 
The keywords include our \textit{email}, \textit{name}, \textit{IMEI}, \textit{serial number}, \textit{Android ID}, \textit{Advertiser ID}, and \textit{Manufacturer}. 
The selection of keywords is based on the personal information that could be involved during our usage of Android apps. Some keywords are mentioned in the official Android documentation~\cite{androidLogRequirement} and existing work~\cite{Xueling2023}.
Additionally, we search for hash values of these data, including MD5 and SHA-1 hashes, as they are considered weak cryptographic algorithms. The personal information is sensitive and should not be included in the logs. If these keywords are found in the logs, it indicates the presence of privacy leaks. 
It is worth noting that our work is not about building privacy-leaking issue detectors for Android app logs. In our study, the privacy-leaking issues we found are 100\% accurate because we use a dataset we constructed, and our privacy information is known.



We also pay particular attention to privacy leaks in logs that originate from third-party Android libraries. Unlike traditional standalone systems, most Android apps require communication with remote servers. If we observe privacy leaks from third-party library logs, it implies that users' private information could be transmitted over the network to the third-party library for further analysis. Such widespread dissemination could have incalculable consequences. 
We determine the source of Android logs by searching for log tags, as they usually reflect the components that generate logs.
Based on the log example below, the keyword (highlighted) represents the source of this log. In this example, the log is generated by a third-party Android library, Localytics~\cite{Upland2024}.


\noindent
\responsecolorbox[gray!10]{
11-15 10:10:05.958 12287 12497 V {\color{blue}Localytics}: first\_android\_id=a54eccb914c21863
}

\noindent  \textbf{\underline{Results.}} 
After searching, we did find that privacy leaks are common in Android logs. We identified 610 privacy leak points in the old dataset and 651 in the recent dataset. Detailed information about privacy leaks, including information content, information types, and quantities, can be found in Table~\ref{tab:table_rq2_info}. According to the table, the most common type of leaked privacy information is the manufacturer (54.10\% in the old dataset and 86.34\%, which dominates the privacy leak cases. Android ID is common in privacy leak cases as well (25.90\% in the old dataset) but it no longer appears in the recent dataset, suggesting that developers have become more cautious in handling or logging Android ID in newer app versions.
There are also many cases of email leaks (15.41\% in the old dataset and 1.08\%, which should draw attention from Android users. 
Notably, the leakage of user names increases substantially in the recent dataset (from 3 in the old dataset to 82 in the recent dataset), indicating a rising risk of exposing personally identifiable information in newer app versions.
The remaining privacy leak cases are relatively few, all less than 10, but users should be cautious about their information being transformed into other forms, such as hash values. Especially for Android ID, it is the most susceptible to being transformed into hash values and exposed to the public.


Different types of information vary in the level of severity they pose to users when leaked. In our results, most privacy leaking instances come from device manufacturers (54.10\%). Since this information cannot be used by malicious attackers to precisely identify users, leaks involving manufacturers are often overlooked. However, such information still needs to be protected. For example, e-commerce sites may adjust their pricing based on your devices~\cite{Wilson2014}. Additionally, other privacy leaking instances we report involve information that is highly specific to individuals and represent more serious cases of privacy leakage.



\begin{table}[ht]
\small
\centering
\caption{A summary of privacy leak cases in Android Logs.} \label{tab:table_rq2_info}    
\begin{threeparttable}
\begin{tabular}{@{}llrlr@{}}
\toprule
 \multirow{2.7}{*}{\textbf{Info. Type}}& \multicolumn{2}{c}{\textbf{Old Dataset}} & \multicolumn{2}{c}{\textbf{New Dataset}} \\
\cmidrule(l{1pt}r{1pt}){2-3}
\cmidrule(l{1pt}r{1pt}){4-5}
 & \textbf{Info.} & \textbf{Count} & \textbf{Info. } & \textbf{Count} \\
\midrule
Manufacturer	    & motorola	&  330  & Pixel 5a  & 562 \\ 
Android ID        & a54e...     &  158  & 1169...   & 0\\
Email           & ...@gmail.com &	 94   & ...@gmail.com & 7\\
AndroidID (SHA-1)	& c5ad...     &  8    & 4b7a...   & 0\\
Serial	        & ZX1G...     &  7   & 1703...   & 0\\
Android ID (MD5)  & 7a43...     &  6    & 8f0d...   & 0\\
Serial (MD5)      & 4ebb...     &  4    & c659...   & 0\\
Name	            & ...        &  3  & ...   & 82\\
\midrule
\textbf{Total}		&                   & 610   &   & 651\\                
\bottomrule
\end{tabular}
\end{threeparttable}
\end{table}

After studying the distributions of privacy leaking instances, we further examined the origins of these logs--whether they originated from the Android app internal calls or third-party library calls--by extracting log tags. Table~\ref{tab:table_third_party} presents our study results. Out of a total of 610 privacy leaking points, we identified 4.10\% privacy leaks originating from Android third-party libraries, i.e., the majority of privacy leak instances stem from the Android app internal calls, which implies that Android apps are highly likely to leak privacy on their own. 
Additionally, privacy leaking logs from third-party libraries demonstrate there is a chance that they may transmit users' personal information to Android third-party libraries.
In these instances of privacy leaks from third-party libraries, Localytics~\cite{Upland2024} is the most common Android third-party library that leads to privacy leaks. We have also identified a few instances of privacy leaks in Unity Ads~\cite{Unity2024} and MoPub~\cite{MoPub2026}. We strongly recommend developers of these third-party libraries to enhance the protection of user privacy.

\begin{table}[ht]
\small
\centering
\caption{A summary of privacy leak cases regarding third-party libraries in Android Logs.} \label{tab:table_third_party}    
\begin{threeparttable}
\begin{tabular}{@{}lrlr@{}}
\toprule
\multicolumn{2}{c}{\textbf{Old Dataset}} & \multicolumn{2}{c}{\textbf{New Dataset}} \\
\cmidrule(l{1pt}r{1pt}){1-2}
\cmidrule(l{1pt}r{1pt}){3-4}
\textbf{Third-Party Lib.} & \textbf{Count} &\textbf{Third-Party Lib.} & \textbf{Count}\\

\midrule
Localytics~\cite{Upland2024}  & 15 &Braze~\cite{Braze2026}     & 11 \\ 
Unity Ads~\cite{Unity2024}	  & 6 & AppsFlyer~\cite{Appsflyer2026}    & 4  \\
MoPub~\cite{MoPub2026}	      & 4  & OkHttp~\cite{Okhttp2026}   & 3 \\
/& /& Applovin~\cite{Applovin2026}    & 1 \\
\midrule
\textbf{Total}                 & 25 & & 19 \\
\bottomrule
\end{tabular}
\end{threeparttable}
\end{table}

From the app perspective, privacy leak is a very common issue. Among the 83 Android apps we analyzed, 35 had privacy leaking issues.
Among the new dataset, we found privacy leakage issues in 51 out of 78 Android apps.
Additionally, we found that most Android apps have a small number of privacy leaking issues, with only a few privacy leaking cases per log file. Table~\ref{tab:table_apps} and \ref{tab:table_apps2} list all Android apps with privacy leak cases over ten. This imbalance suggests that privacy leaks are not easily detectable for most Android apps, as it is difficult for developers to manually identify small amounts of leaked personal information from thousands of log entries. Therefore, developers may potentially overlook this issue. 
An interesting example is the Paper Toss app itself, which does not generate any privacy-leaking logs; all privacy-leaking instances originate from third-party libraries. Therefore, when invoking third-party libraries, Android apps should check whether they transmit user privacy to third-party libraries and carefully select third-party libraries with which they want to communicate.

\begin{table}[ht]
\small
\centering
\caption{Top-7 Android apps with the highest leaking issues (old dataset).} \label{tab:table_apps}    
\begin{threeparttable}
\begin{tabular}{@{}lrr@{}}
\toprule

\textbf{Android App} & \textbf{Third-Party Lib.} &\textbf{Count} \\
\midrule
Capital One Mobile & UnityAds (2), MoPub (2)	& 169  \\
Lime Player	& / &92   \\
GoNoodle	& / &90   \\
Weather	    & / &78    \\
Go Weather	& MoPub (2) & 67 \\
Postmates	& / &28   \\
Paper Toss	& Localytics (15) &15       \\
\midrule
\textbf{Total}	& 21   & 539 \\                
\bottomrule
\end{tabular}
\end{threeparttable}
\end{table}



\begin{table}[ht]
\small
\centering
\caption{Top-7 Android apps with the highest leaking issues (new dataset).} \label{tab:table_apps2}    
\begin{threeparttable}
\begin{tabular}{@{}lrr@{}}
\toprule

\textbf{Android App} & \textbf{Third-Party Lib.} &\textbf{Count} \\
\midrule
Amazon Shopping & / & 314  \\
Prime Video	& / & 41   \\
1Weather Forecasts \& Radar	& / & 24   \\
Cash App	    & / & 20    \\
FOX Sports	& Braze (11) & 20 \\
Pixel Art	& Appsflyer(1) & 19 \\
Tinder	& / & 18 \\

\midrule
\textbf{Total}	& 12   & 456 \\                
\bottomrule
\end{tabular}
\end{threeparttable}
\end{table}

Through this RQ, we found that despite Android official documents~\cite{Android_doc_2024, androidLogRequirement} emphasizing the importance of privacy and security in logs and highlighting that logs should not include PII, we can still find privacy information in Android app logs. Developers have not fully followed the Android official documents.

\begin{Summary}{Summary of RQ2}{}
Privacy leaks in Android app logs are common. Most privacy leaks originate from Android app internal calls, but Android apps may transmit user information to Android third-party libraries, leading to privacy leaks.
\end{Summary}
\vspace{0.1cm}

\subsection{Characteristics of Privacy Leaking Logs (RQ3).}

\noindent \textbf{\underline{Approach.}} 
To study the characteristics of Android app logs that could potentially leak users' privacy information, we analyze two dimensions: (1) \textit{Logging level}. Logging levels indicate the degree of importance developers place on logs. Logging levels have a ranking, where logs with high-ranking logging levels could be printed during software runtime, and vice versa. According to the Android logs documentation~\citet{Android2024}, there are five commonly used logging levels (from highest to lowest): \inline{ERROR}, \inline{WARN}, \inline{INFO}, \inline{DEBUG}, and \inline{VERBOSE}. In Android logs, they are respectively marked by uppercase initials; for example, `E' stands for \inline{ERROR}.
Logs with logging levels (\inline{ERROR}, \inline{WARN}, and \inline{INFO}) equal to or greater than \inline{INFO} should always be printed during software runtime, while logs with logging levels less than \inline{INFO} could be omitted. This also implies that privacy leaking issues in logs with logging levels greater than or equal to \inline{INFO} may result in greater consequences because such logs will always be printed during software runtime, whereas privacy leaking issues in those with logging levels less than \inline{INFO} may result in minor consequences since they are not recommended to be printed. \inline{DEBUG} logs are compiled but removed during software runtime, while \inline{VERBOSE} logs are completely ignored by the software system. To study logging level distribution, we allowed all logs to be printed during software runtime.
(2) \textit{The location of privacy leaks in logs}. We categorize the log content to observe which areas within the Android app logs are more prone to leaking user privacy. 
This categorization was accomplished through a manual study. Each Android app log with privacy leaks was independently labeled by two participants, followed by a discussion to reach a final agreement if disagreements exist.

\begin{figure*}%
    \centering
    \subfloat[\centering Old Dataset]{{\includegraphics[width=0.44\textwidth]{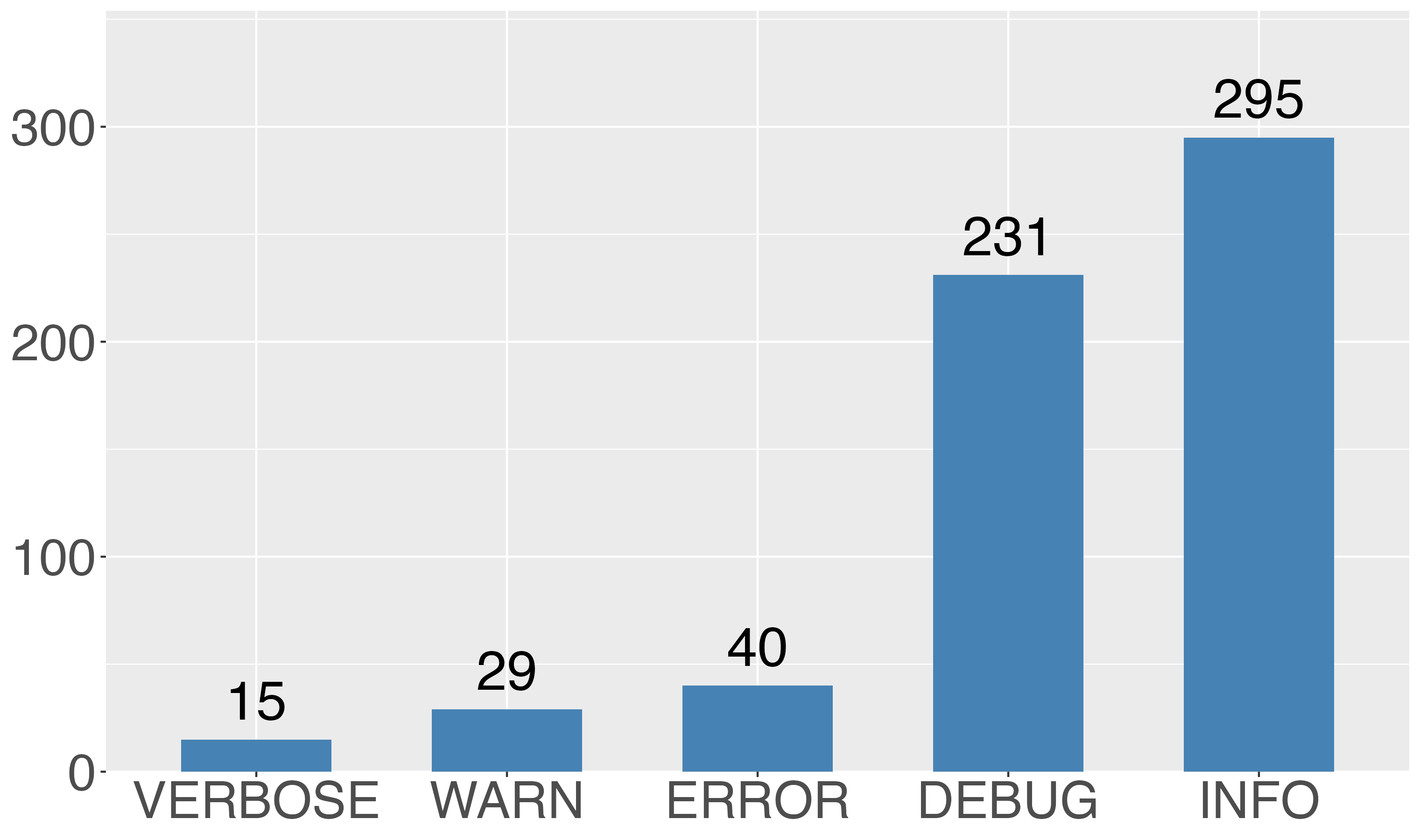} }}%
    \subfloat[\centering New Dataset]{{\includegraphics[width=0.3\textwidth]{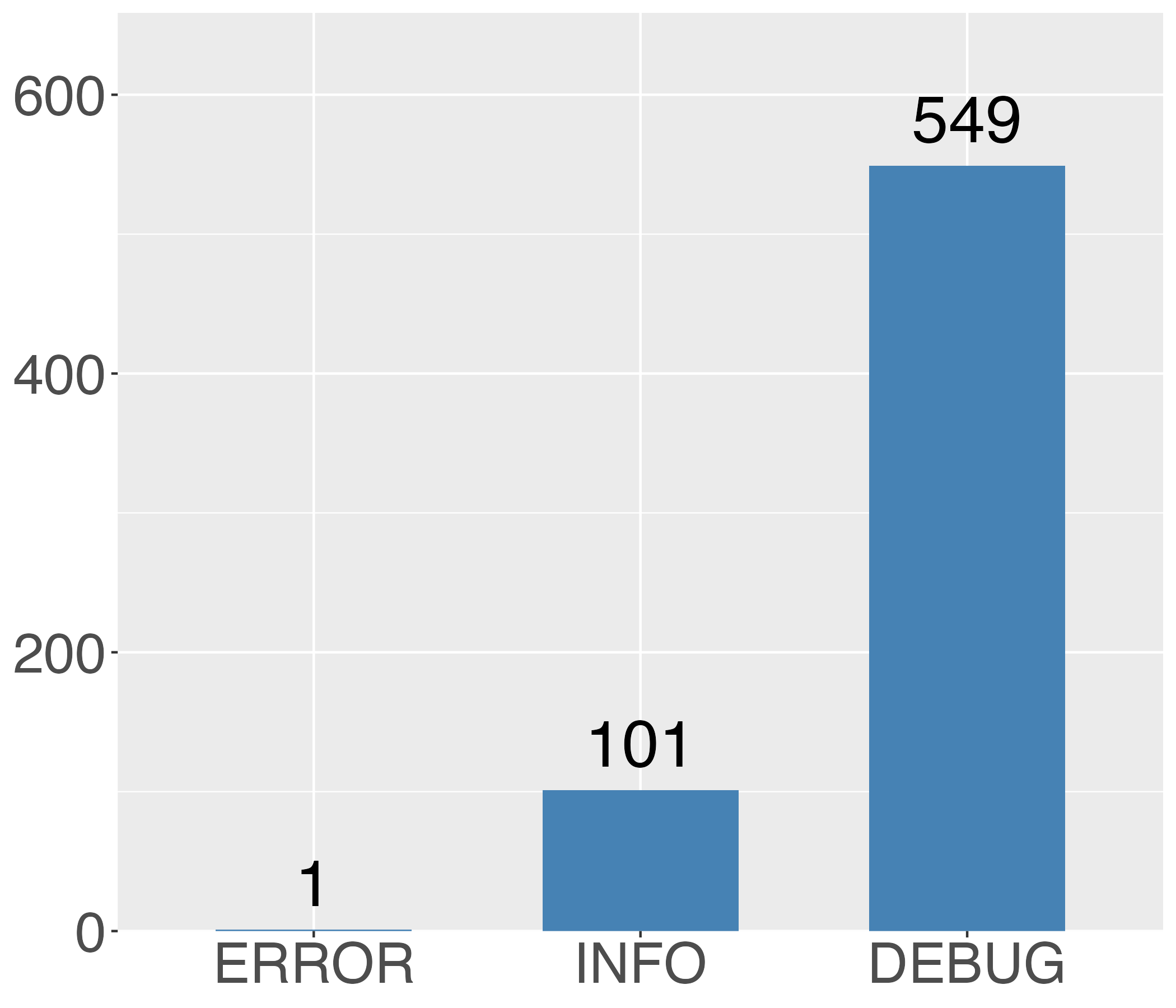} }}%
    \caption{Logging level distribution for the android logs with privacy leaks.}%
    \label{fig:log_levels}%
    \vspace{-1.5em}
\end{figure*}

\noindent  \textbf{\underline{Results.}}
Figure~\ref{fig:log_levels} presents the majority of logs (59.67\%) in the old dataset and a considerable fraction in the new dataset (15.67\%) were generated by logging statements with higher logging levels ($\geq$ \inline{INFO}), which indicates that a significant amount of privacy leaks could occur during Android app runtime. Despite the prevalence of these privacy leaks, it is fortunate that a non-trivial amount of user privacy leaks, although involved in program compilation, ultimately get omitted during software runtime (\inline{DEBUG} logs). 



\begin{figure*}%
    \centering
    \subfloat[\centering Old Dataset]{{\includegraphics[width=0.44\textwidth]{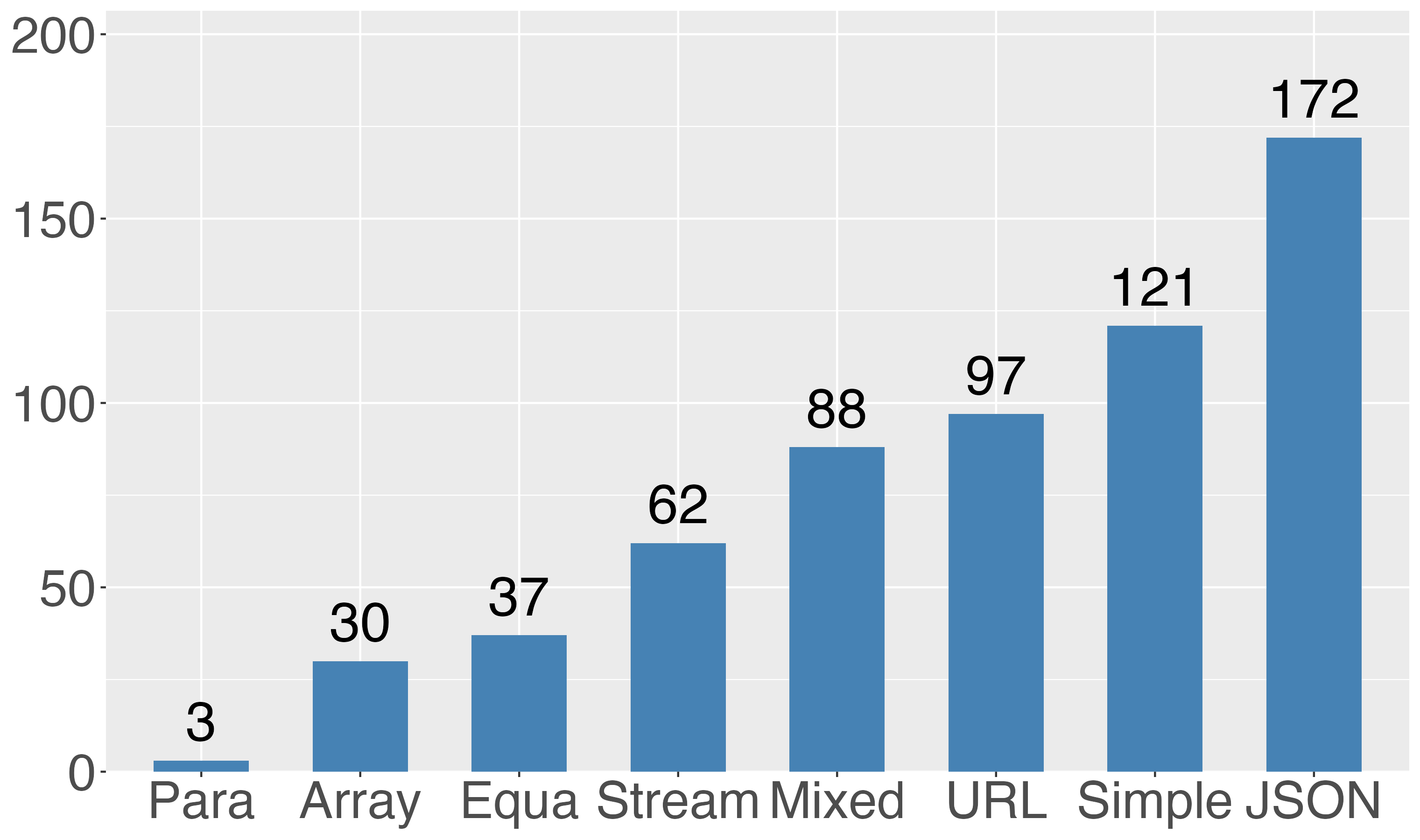} }}%
    \subfloat[\centering New Dataset]{{\includegraphics[width=0.31\textwidth]{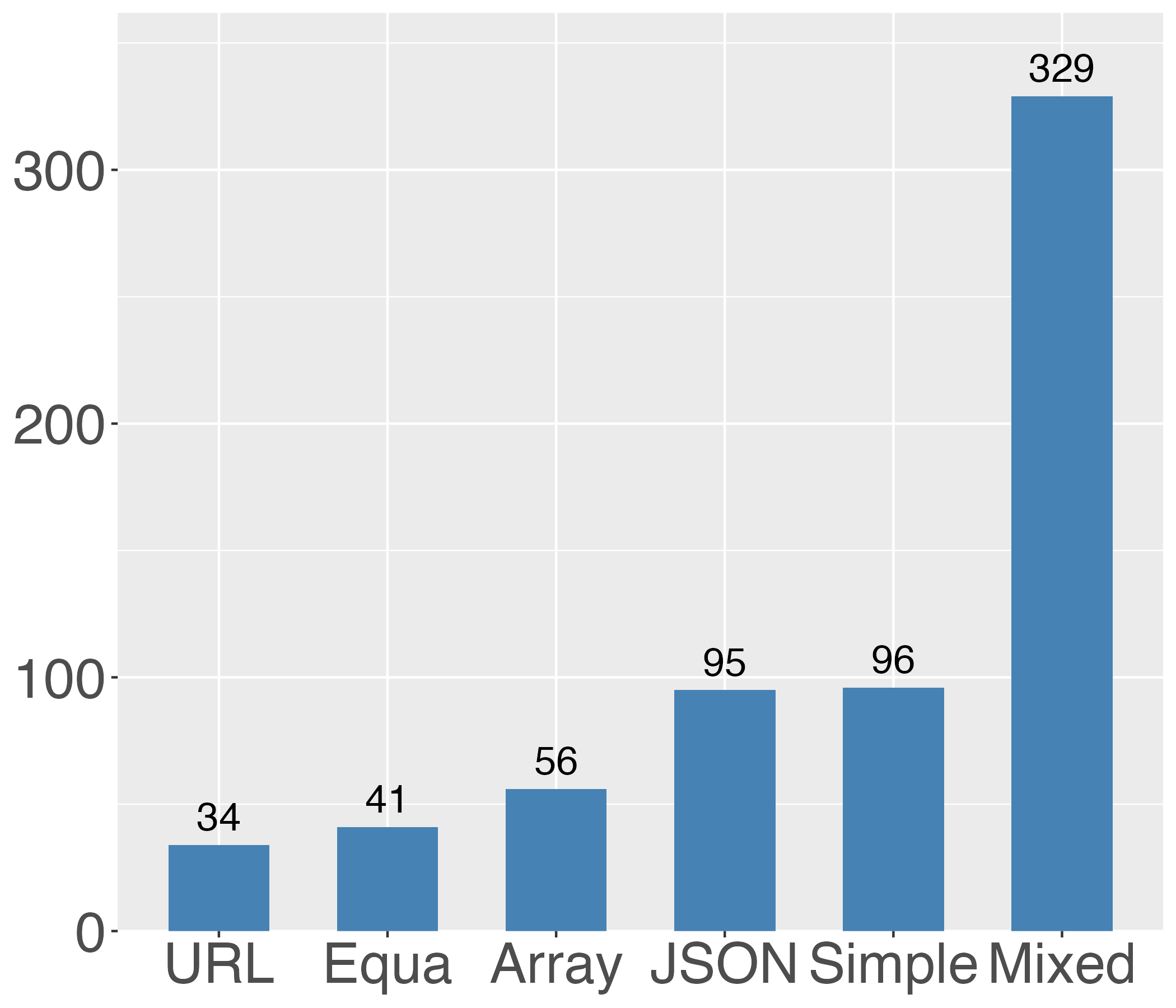} }}%
    \caption{The location of privacy leaks in the Android app logs.}%
    \label{fig:log_loc}%
    \vspace{-1.5em}
\end{figure*}

According to the analysis of logging level distribution, we can infer that privacy leak in Android app logs is a significant issue, as a non-negligible portion of leaked user privacy information ultimately gets printed and stored in the logs. In addition, we also investigated the location where users' leaked privacy is stored in the Android app logs to analyze users' awareness of privacy leaks and the challenges of mitigating this issue. The study results indicate that leaked user privacy can be found in eight different locations within the Android app logs as shown in Figure~\ref{fig:log_loc}.



\paragraph{\underline{Privacy information embedded within JSON (28.20\% vs 14.59\%)}}  In this category, users' privacy is stored within JSON data. Privacy information embedded within JSON structures accounts for 28.20\% of cases in the old dataset, compared to only 14.59\% in the new dataset.
In the example below, the leaked user privacy is the user's name, highlighted in red. When developers need to log complex data structures such as JSON, they usually have a vague understanding of the overall data structure. 
It should be noted that when developers create logging statements to record complex data structures, they often do not write a static string in the logging statements, but rather use variables, for example, including a variable in the logging code to store JSON data. This is why the details in the complex data structures could be hidden from developers. Therefore, developers may overlook checking the content of complex data, which has the potential to leak privacy. 
The presence of extensive JSON data structures in Android app logs complicates users' efforts to protect their privacy.

\noindent
\responsecolorbox[gray!10]{
09-26 15:54:09.296 20811 21230 D OkHttp : \textcolor{blue}{\{``status":\{``code":0,``message":\\``OK"\},``data":\{``userInfo":\{``id":``5d8d256f0a46770001e9d28c",``name":``\textcolor{red}{U...}",\\``avatar":``https://cw-lens.dailyinnovation.biz/paintByNumber/userAvatar/9cd\\a60791ae52cb4e1529c9a1ae7b563"\},\\``lastSync":1569531249\}\}}
}

In addition to strictly adhering to JSON-formatted data, this category also includes a few JSON-like data structures, such as the example below, where key-value pairs are not enclosed in parentheses.

\noindent
\responsecolorbox[gray!10]{
09-26 16:40:16.512 588 619 D MSDG[SmartSwitch]ManagerHost: \textcolor{blue}{Vendor : \textcolor{red}{motorola} Type:user isRightAPK:false SDK:[25]}
}

\paragraph{\underline{Privacy information as a single token (19.84\% vs 14.75\%)}} Privacy information leaks are stored within a single token (a word). This is the most straightforward method to demonstrate leaked user privacy without involving any complex data structures. In such cases, developers may be aware of the information they need to print since this information comes from predefined text or a simple variable in logging statements, i.e., developers may have a certain anticipation of the information that needs to be printed. 
The example below presents such a case of user privacy leak in Android app logs. ``c5ad...'' represents the \inline{SHA-1} hash of the Android ID. To protect privacy, we only display the first four digits. 
In the old dataset, we found 610 leaked privacy cases, only 19.84\% of examples belong to this category, compared with 14.57\% in the new dataset. This implies that the majority of leaked privacy information is included in more complex data structures, posing challenges for developers who may be unaware of the information being leaked.

\noindent
\responsecolorbox[gray!10]{
09-26 18:13:52.277 19836 19902 E BCookieProvider: \textcolor{red}{c5ad...}}


\begin{table}[ht]
\small
\centering
\caption{Detailed categorization for leaked privacy information as a complete token within Android app logs.} \label{tab:table_complete_token}    
\begin{threeparttable}
\begin{tabular}{@{}lrr@{}}
\toprule

\textbf{Log category} & \textbf{Old Dataset} & \textbf{New Dataset}\\
\midrule
Parameter information       &  43   & 8 \\
Job event                   &  42   & 0 \\
Error message	            &  24   & 0 \\
Android ID                  &   8   & 0 \\
User's login status         & 4     & 1 \\
System information          &  0 &  87  \\
\midrule
\textbf{Total}		      & 121     & 96\\                
\bottomrule
\end{tabular}
\end{threeparttable}
\end{table}

To explore further into this category and understand developers' perception of privacy leaks and the challenges they face, we have further refined this classification. Table~\ref{tab:table_complete_token} provides a detailed breakdown, which encompasses five log categories:

\begin{itemize}
    \item \textbf{Parameter information}. These logs all originate from the same event template (see examples below). Due to the lack of restrictions and descriptions for the parameters to be printed by this template, it results in the printing of users' private information. For such logs, we urge developers to carefully consider whether information needs to be printed rather than indiscriminately printing parameter information.

\noindent
\responsecolorbox[gray!10]{
09-05 19:47:13.982 17761 17836 I Return string detection printStackTrace and parameter:: \textcolor{red}{motorola}\\
09-05 19:45:59.680 17761 17844 I Return string detection printStackTrace and parameter:: \textcolor{red}{4ebb...}\\
09-05 19:45:59.680 17761 17844 I Return string detection printStackTrace and parameter:: \textcolor{red}{ZX1G...} }

    \item \textbf{Job events} indicates that developers print job event information in the software logs. In this category, while the content of the software logs may vary, they all contain similar job event information generated from the same log template. In the following example, the blue section represents job event information, which is included in all logs of this category (with some individual details differing, such as job ID values). It is obvious that the leaked user email was not manually inserted by developers into logging statements but rather originated from this job event template. Therefore, when developers need to print job event information, especially when using templates, they should be cautious about whether the templates might inadvertently expose their private information.

\noindent
\responsecolorbox[gray!10]{
09-14 15:51:24.170 2171 2183 D SyncManager: failed sync operation \textcolor{blue}{JobId: 107737, \textcolor{red}{...@gmail.com} u0 (com.google), gmail-ls, SERVER, reason: 10019}, SyncResult: databaseError: true stats [] }

    \item \textbf{Error messages}. These logs are generated when an Android app encounters an error and contain information related to that error. Privacy leaks often stem from developers intending to print error messages associated with user private information, particularly during user authentication processes. In the following example, the program attempts to fetch an authentication token for a user account but fails, which results in the user's account being printed in logs. Developers should exercise extra caution when handling events involving user privacy, particularly during user identity authentication, to avoid inadvertently leaking user privacy.

\noindent
\responsecolorbox[gray!10]{
10-30 20:20:47.374 7881 21069 E HeterodyneSyncTaskChime: Failed getting auth token: for \textcolor{red}{...@gmail.com} [CONTEXT service\_id=51 ] }

\item \textbf{Android ID}. This category of logs is similar to the example shown above in Table~\ref{tab:table_complete_token}. In this log category, 
developers specify only the \inline{SHA-1} hash of the Android ID to be logged, while the rest of the content is for log metadata. 
This direct printing of user privacy should be avoided.

\item \textbf{User's login status.} 
Developers may leak user privacy by using logs to print user login status information. For example, in the following log, developers intend to print the user's account name. While it is understandable that developers may want to track user login situations, they should avoid printing data related to user privacy.

\noindent
\responsecolorbox[gray!10]{
10-30 15:07:13.413 9680 9680 I Indeed/GoogleAuthManage: google login as \textcolor{red}{...@gmail.com} }

\item \textbf{System information}
Logs in this category primarily record system or device runtime information for debugging purposes, such as device models and user name. As shown below, while the first log records device information, the second log exposes the user's name, which may lead to privacy leaks. 

\noindent
\responsecolorbox[gray!10]{
01-28 09:35:50.997 I/org.webrtc.Logging(23070): WebRtcAudioManager: \textcolor{red}{Pixel 5a} is blacklisted for OpenSL ES usage!\\
01-28 09:27:04.516 D/com.amazon.mShop.android.shopping/LocalDataStor\\ageEncryptor(13851): Data after decryption is \textcolor{red}{Kirito}
}

\end{itemize}

According to the analysis above, we can infer that although developers may intentionally place privacy information as a complete token, most instances of such occurrences (70.25\%) actually result from using log templates that inadvertently lead to privacy leaks. Additionally, some instances arise when printing error messages, especially when events related to user identity authentication fail which results in privacy leaks. Only around 9.92\% of privacy-leaking logs in this category are likely intentionally generated by developers.
Moreover, in the new dataset, we identified the system information category, which represents 90.63\% in this group.

\paragraph{\underline{Privacy information embedded within URLs (15.90\% vs. 5.22\%)}}

In this category, users' leaked privacy becomes part of the URL. In the example below, the user's device manufacturer is printed as a parameter value in the URL. To save space, we have omitted some parameter information in the middle of the URL. 

\noindent
\responsecolorbox[gray!10]{
09-26 18:21:10.572 23497 23573 D MoPub : Loading url: https://ads.mopub.com/m/ad?v=6\&...\&dn=\textcolor{red}{motorola}...
}

Additionally, we have found instances of privacy leaks originating from Android package names, which we also include in this category due to their similarity to URLs. In the example below, the log attempts to print the package name, which contains information about the device manufacturer.

\noindent
\responsecolorbox[gray!10]{
10-30 16:33:11.394 954 17890 I PFTBT : Initiating full-data transport backup of com.\textcolor{red}{motorola}.entitlement
}

\paragraph{\underline{Privacy information embedded within data stream (10.16\%)}}
In this category, logs contain a special data stream delimited by ``$\Vert$'' to separate different tokens. This unique data structure is used to store tabular data. According to the example below, this data stream represents a record in the database. Printing the entire data record without considering the sensitivity of its content can potentially lead to privacy leaks.


\noindent
\responsecolorbox[gray!10]{
11-15 10:19:09.596 14479 14550 D StatisticsManager: Insert static Data to DB:24$\Vert$2019-11-16 00:19:09$\Vert$\textcolor{red}{a54e...}$\Vert$1573791844988a54e\\ccb914c21863$\Vert$US$\Vert$200$\Vert$307$\Vert$6.154$\Vert$184$\Vert$user\_type$\Vert$1$\Vert$0
}

\paragraph{\underline{Privacy information embedded within equations (6.07\% vs 6.30\%)}}
In addition to JSON structures, Android app logs also utilize equations to store key-value pair data. This JSON-like structure often contains a wealth of information that may not be carefully scrutinized by the developers, which leads to privacy leaks.

\noindent
\responsecolorbox[gray!10]{
09-12 18:32:09.471 8267 8267 D BuaReceiver: ====== got intent:Intent { act=android.intent.action.PACKAGE\_ADDED dat=package:com.foxsports.android flg=0x4000010 cmp=com.\textcolor{red}{motorola}.android.buacontactadapter/.BuaReceiver (has extras) }
}

\paragraph{\underline{Privacy information embedded within arrays (4.92\% vs 8.60\%)}} In this category, users' privacy is stored within an array.

\noindent
\responsecolorbox[gray!10]{
09-14 17:30:35.694 29105 29105 I ExoPlayerImpl: Init 7f658fc [ExoPlayerLib/2.10.2] [shamu, Nexus 6, \textcolor{red}{motorola}, 25]
}

\paragraph{\underline{Privacy information as a method parameter (0.49\%)}}
We have identified three instances where users intended to print method signatures, but users' private information was printed as parameter values, leading to privacy leaks.

\noindent
\responsecolorbox[gray!10]{
10-30 16:17:34.155 15624 15836 E p : downloadCorpusUpdates(\textcolor{red}{...@gmail.com}, PEOPLE\_API, agsa) failed
}

In addition to storing users' personal privacy within a single data structure, privacy information can also be stored in various hybrid data structures (14.43\%). Any of the mentioned data structures can be combined with another, with the most common being the mixing of URLs and equations, as shown in the following example.
In contrast, hybrid structures account for 50.54\% of cases in the new dataset, showing a substantial increase compared to the old dataset.

\noindent
\responsecolorbox[gray!10]{
11-15 10:50:19.657 10448 10477 I Babel\_SMS: MmsConfig.loadDeviceMmsSettings from API: mUserAgent=nexus6, mUaProfUrl=http://uaprof.\textcolor{red}{motorola}.com/phoneconfig/nexus6/Profile/nexus6.rdf}

According to the analysis above, we can observe that most leaked privacy information is stored within complex data structures, making it difficult for developers to be aware of their privacy leaks. 
Furthermore, our analysis of privacy leaks in simple data structures (tokens) indicates that developers are usually unaware of privacy leaks as well, with the majority being inadvertent privacy leaks caused by the use of log templates. 
A study~\cite{Yang2013} indicates that
the transfer of sensitive data mentioned does not necessarily imply privacy leaks but rather, user intent should be considered a better indicator to determine whether privacy leaks occur in Android apps. Regardless of whether we study the privacy leaking issues in Android app logs from the perspectives of privacy information existence or user intent, we can conclude that there is indeed a non-trivial privacy leaking issue present in Android app logs.

\begin{figure}[h]
    \centering
    \includegraphics[width=0.65\textwidth]{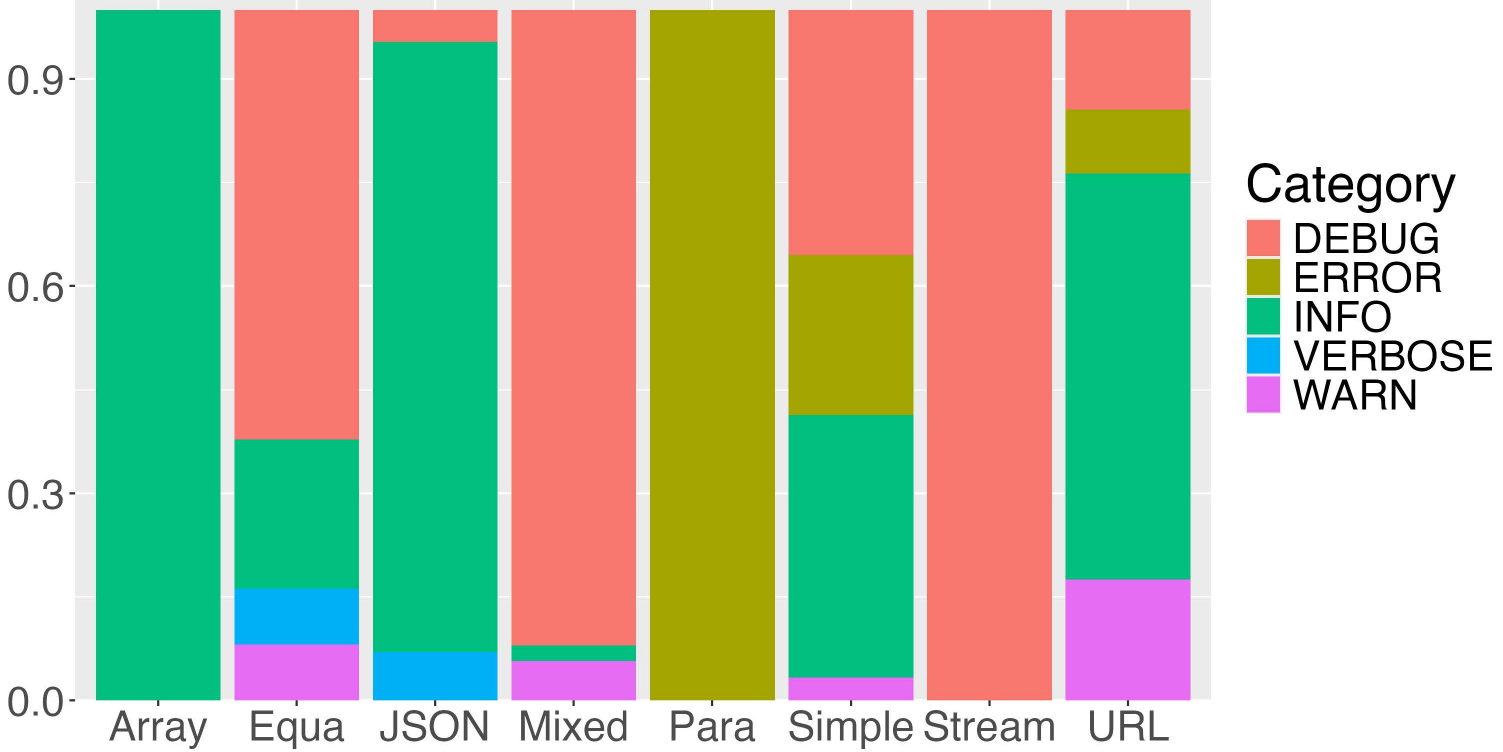}
	\caption{Privacy leaking Android app logs' logging level distribution across different privacy leaking log data structure categories.} 
	\label{fig:log_levels_data} 
 \end{figure}

To understand the relationship between logging levels and different log data structures for privacy leaking logs, we conduct a statistical analysis on the old dataset, and its result is shown in Figure~\ref{fig:log_levels_data}. Within privacy-leaking logs, the distribution of logging levels across different data structures is highly unbalanced. Some categories exhibit only one logging level, such as the \inline{ERROR} level for category \textit{parameters}, \inline{INFO} level for  \textit{arrays}, and \inline{DEBUG} level for  \textit{data streams}. Logs within the same data type category often share similar event templates or even identical log content, which indicates that these logs may originate from either a single logging statement or duplicate logging statements (an existing study indicates the prevalence of duplicate logging statements in software systems~\cite{Li2022}). This insight provides avenues for addressing privacy leaks in Android apps. Specifically, by capturing log templates as indicators to identify relevant logging code, thereby potentially eliminating a majority of privacy-leaking cases in Android app logs.

Data structure categories dominated by logging levels equal to or higher than \inline{INFO} warrant particular attention, including  \textit{parameters},  \textit{arrays},  \textit{URLs},  \textit{simple data}, and  \textit{JSON}. In particular,  \textit{JSON} category demands special attention as it accounts for the majority of all privacy-leaking cases, and 88.37\% of them require printing during app runtime. It is worth noting that since this category does not contain any \inline{WARN} or \inline{ERROR} messages, which are the main focus of developers during software development, privacy leaks within JSON structures may be easily unnoticed by developers.

\vspace{0.1cm}
\begin{Summary}{Summary of RQ3}{}
Developers are usually unaware of privacy leaks in Android app logs. Developers should pay close attention to privacy leaking issues when writing Android app logging code, especially if they intend to print complex data structures.
\end{Summary}
\vspace{0.1cm}

\section{Discussion}\label{sec:discussion}
In this section, we discuss the implications of our study. 

\subsection{Developers' Concerns Correspond to Privacy Leaking Cases in the Android App Logs.}~\label{sec:concern}

In RQ1, when discussing sensitive information replacement strategies, the developers mentioned that HTTP requests and JSON could potentially leak privacy, which aligns with our observations in Android app logs. 
However, it is worth noting that although JSON and HTTP contribute to a significant number of privacy leaking cases in Android app logs, related Q\&A posts are very limited, with only one post each. This indicates that real-world developers' awareness of privacy protection still needs improvement.

Unlike conventional software logs, JSON is not common content in general software logs. We examined a well-known log dataset~\cite{loghub} comprising 2K log lines from each of the 16 popular projects and found only one log line containing JSON data. While the situation is reversed in Android app logs, this characteristic might not be known by developers and has not been brought to their attention.

In RQ1, we noticed a category where developers did not want to protect privacy. 
This category also gives us insight that there may be cases where developers are indifferent to personal privacy being exposed. Within the Android app logs we analyzed, there are instances where developers might be aware of privacy leaks.
 The developers' posts and Android app logs demonstrate that, despite protecting personal privacy being common sense, there are still developers who choose not to follow it and actively log private information. This also indicates that the dissemination of privacy awareness needs to be strengthened.

\subsection{What Do Developers Learn from our Study?}

Our study can provide the following recommendations to developers to safeguard their users' personal information from being leaked in Android app logs.

First of all, when developers write logging code, they should carefully scrutinize the logged information. Developers should avoid deliberately log sensitive data. Moreover, developer should exercise caution when printing complex data structures or variables with values that do not consist of single tokens to prevent inadvertent privacy leaks.

Second, while some real-world developers may hold the misconception that using hash algorithms like \inline{MD5} and \inline{SHA-1} can adequately protect user privacy according to our mining of developer Q\&A forums, it is crucial to recognize that both \inline{MD5} and the \inline{SHA-1} family are considered broken~\cite{NIST2024} and should not be relied upon to safeguard sensitive data.


Third, the privacy-leaking logs we found are not entirely independent. Many share the same log templates. This correlation between logs can facilitate privacy leak detection and elimination in Android app logs, providing valuable insights for developers.

\subsection{Privacy Information is Often Embedded.}

In RQ3, we analyzed the location of privacy-leaking cases in Android app logs and we found that privacy information can exist in various data structures, with the majority embedded in complex data structures. 
This finding not only indicates that developers' awareness of privacy leaks needs to be improved but also presents challenges for detecting privacy leaks. Taking the most common privacy-leaking data structure, JSON, as an example, analyzing JSON data requires not only correctly identifying key-value pairs but also the ability to handle high-dimensional data due to the potential presence of nested JSON structures. As we mentioned in Section~\ref{sec:concern}, JSON is uncommon in general log datasets, so analyzing JSON from software logs has not garnered developers' attention, let alone extracting privacy information from JSON. Our paper brings the issue of analyzing complex data structures in Android app logs to developers' attention. 
Hoping that in the future, developers can create robust privacy-leak detectors capable of identifying privacy leaks within complex data structures from Android app logs.

\section{Threats to Validity} \label{sec:threat}
In this section, we discuss the threats to the validity of our study.

\textbf{Construct Validity.} In our initial research plan, we intended to utilize a log parser to analyze Android app logs to determine whether the tokens representing user privacy were static or dynamic. Static tokens would indicate intentional privacy leaks by developers, as they typically are generated by the predefined text in logging statements, while dynamic tokens would need further analysis. We employed the state-of-the-art log parser Drain3\footnote{\url{https://github.com/logpai/Drain3}} to parse all our Android app logs, collecting log templates for all log lines containing privacy leaks. However, we surprisingly found that the results of log parsing were highly inaccurate, rendering them unsuitable for further analysis. The first step in log parsing often involves tokenizing logs. Due to the presence of numerous complex data structures in Android app logs, tokenizing Android app logs presents significant challenges. Consequently, we abandoned the use of a log parser.

\textbf{Internal Validity.} 
This study involves manual analysis. To mitigate bias introduced by manual analysis, each manually analyzed record is independently verified by at least two participants. In cases where a consensus cannot be reached, a third participant is introduced as a mediator. 

\textbf{External Validity.} 
In this study, we only focused on Android app logs without taking into account apps from other mobile OS platforms. This limitation may introduce a bias in our results toward Android. Nevertheless, Android holds the largest global market share for mobile OS systems~\cite{Howarth2024} and offers extensive mobile apps. The analysis of 83 Android apps in our study ensures sufficient diversity, which could influence a significant number of users.
Additionally, our study only tested Android apps on a device and did not consider the impact of different devices. Since software logs are generally affected to a limited degree by hardware, testing on a device can effectively identify privacy leak issues within Android apps.

\section{Conclusion \& Future Work}\label{sec:conc}

In this paper, we conducted an empirical study on Android app logs to investigate their privacy leak situations. Our study encompasses three dimensions: (1) examining real-world developers' concerns regarding privacy issues related to software logs, (2) studying privacy leaks in Android app logs, and (3) analyzing the characteristics of privacy-leaking Android app logs. To accomplish this study, we also constructed a comprehensive Android app log dataset to fill the gap in the absence of open-source Android app logs. 
We further updated the dataset by collecting logs from newer APK versions of the same set of apps, enabling a longitudinal comparison across app updates.
Our study revealed non-trivial privacy leaking issues within Android app logs, with the majority occurring unbeknownst to developers. The developers' lack of privacy awareness might contradict the consensus on privacy protection.
In the future, we aim to explore the privacy paradox in Android app logs, i.e., examining whether developers' privacy awareness aligns with their declarations. 
We will also investigate privacy leaks from logs of malware Android apps, such as those from AndroZoo~\cite{Allix2016}.


\bibliographystyle{elsarticle-harv} 
\bibliography{ref}

\end{document}